\documentclass[a4paper,openany, 12pt]{book}

\usepackage[text={15.5cm,23cm},centering]{geometry}

\usepackage{mathptmx}
\usepackage{graphicx}
\usepackage{chapterbib} 
\usepackage{subfigure}
\usepackage{color}
\usepackage[colorlinks=true, linkcolor=blue]{hyperref}
\usepackage[english]{babel}  

\def\VEV#1{\left\langle #1\right\rangle} 

\title{Physical Cosmology from the 21-cm Line}
\author{Steven R. Furlanetto}

\begin{document}

\setcounter{chapter}{2}

\chapter{Physical Cosmology From the 21-cm Line}

 {\bf Steven R. Furlanetto, University of California, Los Angeles}
 
 Excerpted from {\it The Cosmic 21-cm Revolution: Charting the first billion years of our Universe}, Ed. Andrei Mesinger (Bristol: IOP Publishing Ltd) AAS-IOP ebooks, \url{http://www.iopscience.org/books/aas}

 \vskip 0.5in

\begin{bf}
  We describe how the high-$z$ 21-cm background can be used to improve both our understanding of the fundamental cosmological parameters of our Universe and exotic processes originating in the dark sector. The 21-cm background emerging during the cosmological Dark Ages, the era between hydrogen recombination and the formation of the first luminous sources (likely at $z \sim 30$), is difficult to measure but provides several powerful advantages for these purposes: in addition to the lack of astrophysical contamination, it will allow probes of very small scale structure over a very large volume. Additionally, the 21-cm background is sensitive to the thermal state of the intergalactic hydrogen and therefore probes any exotic processes (including, e.g., dark matter scattering or decay and primordial black holes) during that era. After astrophysical sources have formed, cosmological information can be separated from astrophysical effects on the 21-cm background through methods such as redshift space distortions, joint modeling, and by searching for indirect effects on the astrophysical sources themselves.
\end{bf}

\section{Introduction}

The previous chapter has shown how the first galaxies and black holes have enormous implications for the 21-cm background. However, all these astrophysical processes occur within the framework of cosmological structure formation -- a process we would like to probe to understand the fundamental properties of our Universe.   This chapter will examine ways in which the 21-cm background can be used to probe the cosmology. Just as fluctuations in the cosmic microwave background and galaxy distribution can be used as probes of cosmology, so can the H~I distribution at $z > 10$. We shall see that the 21-cm background offers an unparalleled probe of the matter distribution in our Universe \cite{tegmark09}: interferometric measurements can, at least in principle, map the distribution of gas over a wide range in both redshift and physical scale. Moreover, we have seen that the amplitude of the 21-cm signal depends sensitively on the thermal state of the IGM. Although the combination of adiabatic expansion and X-ray heating determines that state in the standard scenario, any  ``exotic" process -- like dark matter decay, scattering between dark matter and baryons, X-rays from primordial black holes, etc. -- that changes this energy balance will leave a signature in the 21-cm background. The low temperature of the hydrogen gas before any astrophysical X-ray background forms means that the 21-cm line is an exceptionally sensitive calorimeter for these processes.

In this chapter, we will review some of these potential cosmological probes. We begin in section \ref{cos-dark-ages} with a discussion of cosmology in the ``Dark Ages" before structure forms -- an era that should be uncontaminated by astrophysics, although it is also extraordinarily hard to observe. Then, in section \ref{cos-astro}, we consider how cosmological information can be extracted from the signal even in the presence of astrophysical processes. Finally, in section \ref{cos-complementary}, we briefly point out that 21-cm measurements can offer strong synergies with other cosmological probes.

\section{Cosmology in the Dark Ages} \label{cos-dark-ages}

\subsection{Setting the Stage: the Standard Cosmological Paradigm} \label{cos-standard}

Before astrophysical sources turn on, the 21-cm background depends on the thermal evolution of the intergalactic medium (IGM) and the earliest stages of structure formation in the Universe. We will therefore first describe these processes in the context of the standard cosmological paradigm. 

\subsubsection{Thermal Evolution}

Let us begin with the thermal evolution.  If it were thermally isolated, the IGM gas would simply cool adiabatically  as the Universe expands. For an ideal gas this cooling rate can be written as $(\gamma-1)(\dot{\rho_b}/\rho_b) \bar{T}_e$, where $\rho_b$ is the baryon density, $\bar{T}_e$ is the electron temperature (equal to the hydrogen temperature in this regime), and $\gamma=5/3$ is the adiabatic index of a mono-atomic gas. For gas at the mean density, the factor $(\dot{\rho}_b/\rho_b)=-3H$ due to the Hubble expansion.  

However, the gas is not actually thermally isolated: free electrons may exchange energy with CMB photons through Compton scattering.  Although cosmological recombination at $z \sim 1100$ results in a \emph{nearly}
neutral universe, a small fraction $\bar{x}_e \sim 10^{-4}$ of electrons are ``frozen out" following the recombination process.  These free electrons scatter off CMB photons and, for a long period, maintain thermal equilibrium with that radiation field. The timescale for Compton cooling is 
\begin{equation}
t_{\rm C} \equiv \left( {8\sigma_T a_{\rm rad} T_\gamma^4\over  3m_ec} \right)^{-1} = 1.2 \times 10^8 \left( {1+z \over 10} \right)^{-4} \ {\rm yr},
\label{eq:tcompton}
\end{equation}
where $T_\gamma \propto (1+z)$ is the background radiation temperature (in this case, the CMB), $\sigma_T$ is the Thomson cross section, $a_{\rm rad}$ is the radiation constant, and $m_e$ is the electron mass.

Including both adiabatic cooling and Compton heating, the temperature evolution of gas at the mean cosmic density,  is therefore described by
\begin{equation}
{d \bar{T}_e\over dt}={\bar{x}_e \over (1 + \bar{x}_e)} \left[ {T_\gamma - \bar{T}_e \over t_{\rm C}(z)} \right] -2H \bar{T}_e.
\label{eq:Compton}
\end{equation}
The first term describes Compton heating. For an electron-proton gas, $\bar{x}_e = n_e/(n_e+n_H)$ where $n_e$ and $n_H$ are the electron and hydrogen densities; the relation is more complicated when helium is included. This prefactor appears because the electrons must share the energy they gain from Compton scattering with the other particles. The last term on the right-hand-side of equation (\ref{eq:Compton}) yields the adiabatic scaling $\bar{T}_e\propto (1+z)^2$ in the absence of Compton scattering. 

\begin{figure} 
\centerline{\includegraphics[height=10cm]{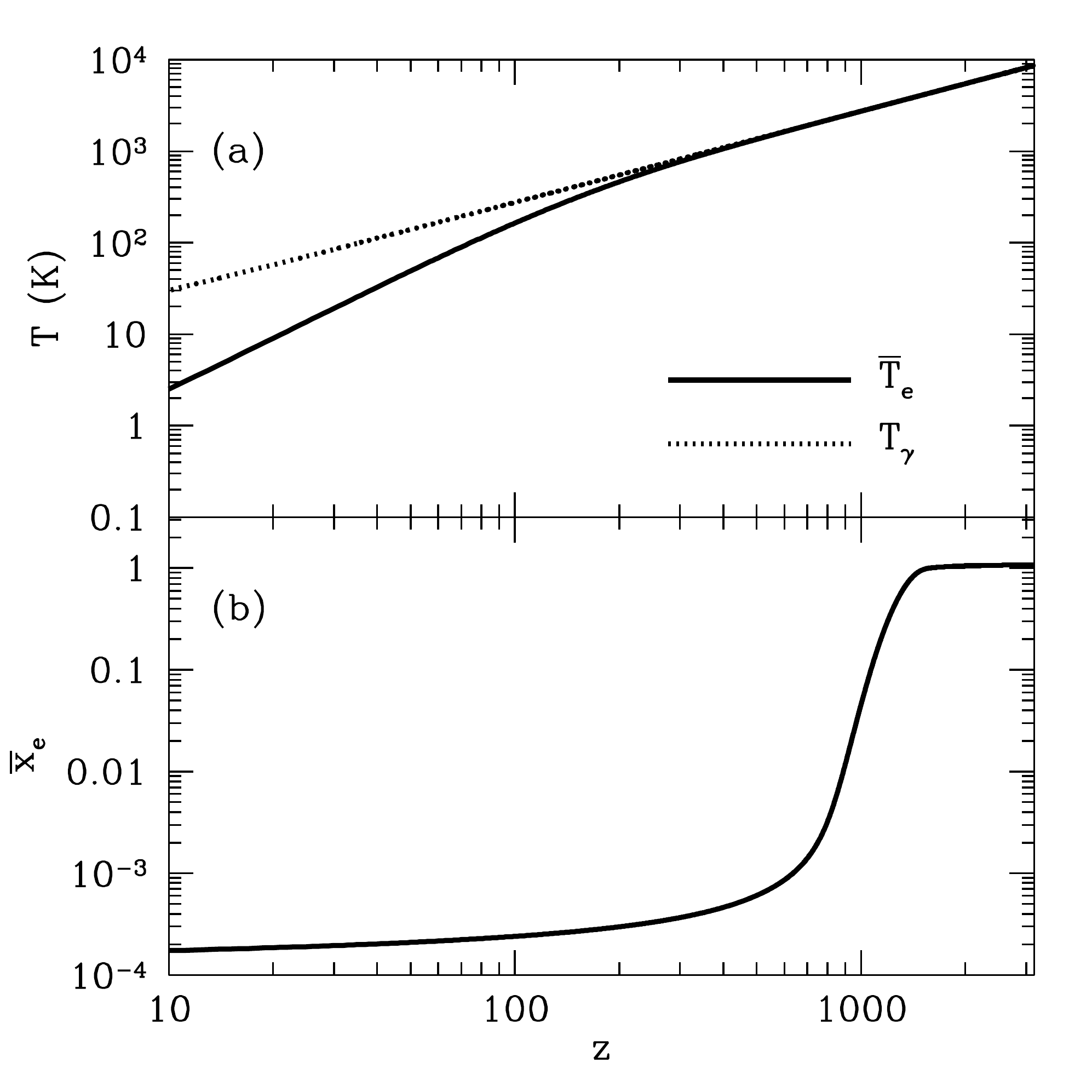}}
\caption{Thermal and ionization history during the Dark Ages (panels a and b, respectively).  The free electron fraction decreases rapidly after recombination at $z \sim 1100$ and then ``freezes-out" at later times.  Meanwhile, Compton scattering keeps $\bar{T}_e \approx T_\gamma$ until $z \sim 200$, after which the declining CMB energy density and small residual ionized fraction are no longer sufficient to maintain thermal equilibrium between the gas and CMB.  Past that point, $\bar{T}_e \propto (1+z)^2$ as appropriate for an adiabatically expanding
non-relativistic gas. These results were produced with the publicly available code RECFAST (http://www.astro.ubc.ca/people/scott/recfast.html). }
\label{fig:darkage-thermal}
\end{figure}

The temperature evolution therefore depends on the residual fraction of free electrons after cosmological recombination \cite{seager99, alihaimoud11, chluba11}.  It is easy to estimate this residual fraction, but in detail depends on the complex physics of hydrogen recombination. In a simple picture, the hydrogen recombination rate is
\begin{equation}
{d \bar{x}_e \over dt} = - \alpha_B(T_e) \bar{x}_e^2 \bar{n}_H
\label{eq:dxedt}
\end{equation} 
where $\alpha_B \propto T_e^{-0.7}$ is the case-B recombination coefficient. (The case-B coefficient ignores recombinations to the ground state, which generate a new ionizing photon and so do not change the net
ionized fraction.)  In the standard cosmology, the fractional change in $\bar{x}_e$ per Hubble time is therefore 
\begin{equation} 
{\dot{n}_e \over H^{-1} n_e} \approx 7 x (1+z)^{0.8},
\end{equation}
where $n_e$ is the electron fraction. Electrons ``freeze-out" and cease to recombine effectively when this factor becomes of order unity; after that point, the Hubble expansion time is shorter than the recombination time.  More precise numerical calculations account for the large photon density during cosmological recombination, line emission, and recombinations to higher energy levels, amongst other factors  \cite{seager99, alihaimoud11, chluba11}. Figure~\ref{fig:darkage-thermal} shows the result of one such calculation, which yields $x \approx 3 \times 10^{-4}$ by $z \approx 200$. 

Inserting this electron density into equation~(\ref{eq:Compton}), we find that the small fraction of residual electrons maintains thermal equilibrium between the gas and CMB down to $z \approx 200$, when Compton heating finally becomes inefficient.  Figure~\ref{fig:darkage-thermal} shows a more exact calculation: note how the gas and CMB temperatures begin to depart at $z \sim 200$, after which the gas follows the expected adiabatic cooling track.

\subsubsection{Density Fluctuations}

Measurements of fluctuations in the 21-cm background depend on variations in the density, spin temperature, and ionization fraction. We therefore briefly consider here how fluctuations in those quantities can emerge during the Dark Ages. A complete treatment of the power spectrum and structure formation is well outside the scope of this work, but we will summarize the key points for understanding the 21-cm signal. 

Density fluctuations grow through gravity and, in the linear regime, their statistical properties can be calculated precisely. Combining the mass and momentum conservation equations for a perfect fluid with the Poisson equation for the gravitational potential yields an evolution equation for the Fourier transform $\delta_{\bf k}$ of the fractional density perturbation $\delta$,
\begin{equation}
\frac{\partial^2\delta_{\bf k}}{\partial t^2}+2 H
\frac{\partial\delta_{\bf k}}{\partial t}=4 \pi G \bar{\rho} \delta - {c_s^2 k^2 \over a^2} \delta_{\bf k}, 
\label{eq:linper} 
\end{equation}
where the last term is the pressure force (which vanishes for cold dark matter) and $c_s^2$ is the sound speed.  This linear equation has two independent solutions, one of which grows in time and eventually comes to dominate the density evolution.\footnote{Note that this solution uses Eulerian perturbation theory, which breaks down when $\delta$ is still relatively small. It suffices during the Dark Ages, but greater accuracy is necessary during later eras. } 

While the fluctuations are small -- and thus very nearly Gaussian -- the density field can be accurately characterized by the power spectrum,
\begin{equation}
P({\bf k}) = \VEV{\delta_{\bf k} \delta^*_{\bf k'}} = (2 \pi)^3 \delta^D({\bf k} -{\bf k}') P({\bf k}). 
\label{eq:pk-defn}
\end{equation}
The power spectrum is the expectation value of the Fourier amplitude at a given wavenumber; for a homogeneous, isotropic universe it depends only on the magnitude of the wavenumber.\footnote{Henceforth we will suppress the ${\bf k}$ subscript for notational simplicity.} The power spectrum therefore represents the variance in the density field as a function of smoothing scale

\begin{figure} 
\centerline{\includegraphics[height=10cm]{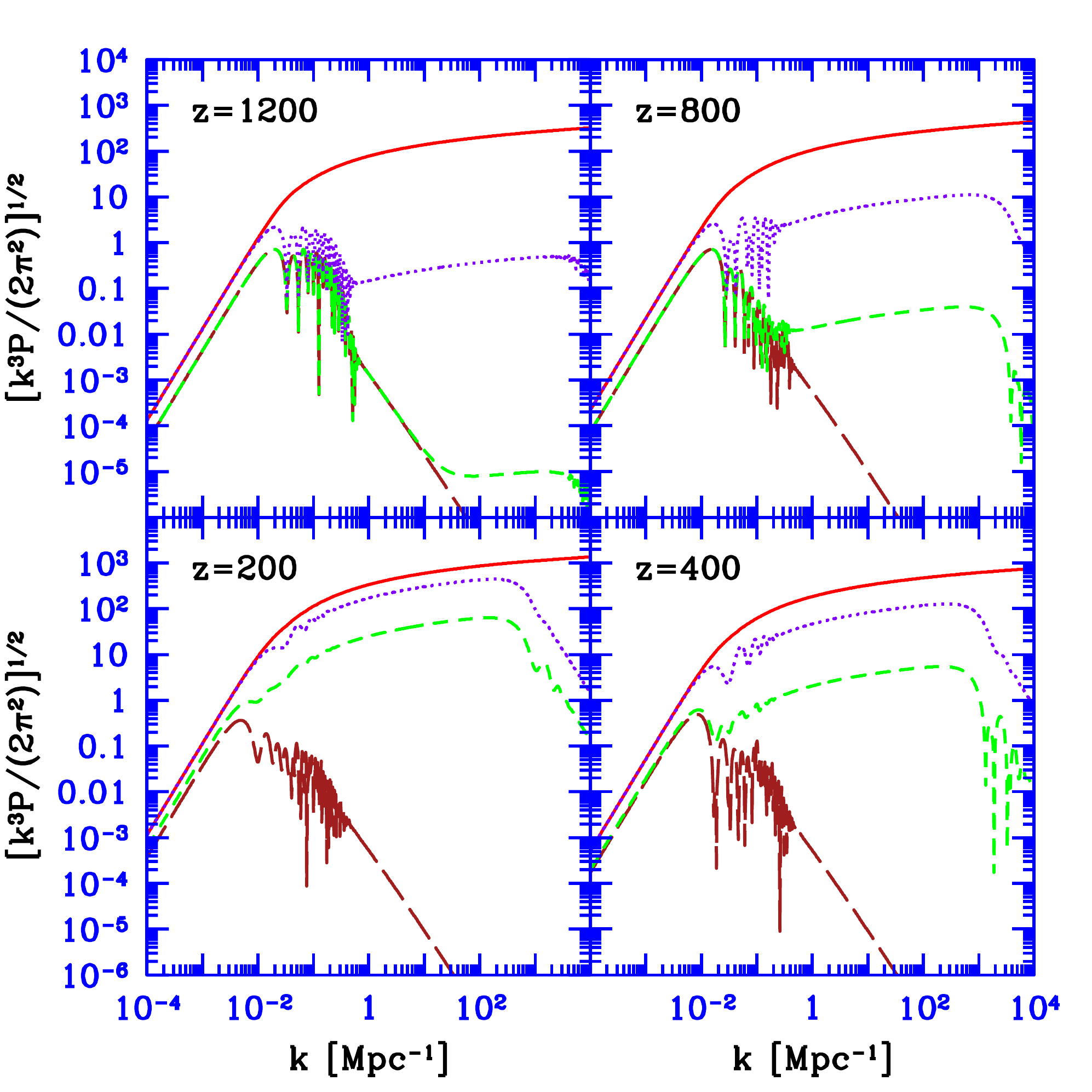}}
\caption{Power spectra for density and temperature fluctuations as a function of comoving wavenumber at four different redshifts. The curves show dimensionless power spectra for the CDM density (solid), baryon density (dotted), baryon temperature (short-dashed), and photon temperature (long-dashed). These curves do not include the relative streaming of the baryons and cold dark matter. Reproduced from Naoz, S. \& Barkana, R., ``Growth of linear perturbations before the era of the first galaxies,Ó \emph{Monthly Notices of the Royal Astronomical Society}, vol. 362, pp. 1047--1053. Copyright OUP 2005.}
\label{fig:nb-pk}
\end{figure}

The solid curve in Figure~\ref{fig:nb-pk} shows the cold dark matter power spectrum at several redshifts during the Dark Ages (taken from \cite{naoz05}). The most obvious feature is the flattening at $k \sim 0.1$~Mpc$^{-1}$, which results from stagnation in the growth of small-scale structure during the radiation era. The power spectrum is otherwise quite simple. However, the 21-cm line will probe the fluctuations in the IGM \emph{gas}, which we will require more physics to understand. 

For example, equation~(\ref{eq:Compton}) describes the evolution of the \emph{mean} IGM temperature.  However, that field actually fluctuates as well, for two reasons \cite{naoz05}.  First, the CMB temperature itself has fluctuations, so each electron will scatter off a different local $T_\gamma$: the power spectrum of the CMB temperature fluctuations is shown by the long-dashed curve in Figure~\ref{fig:nb-pk}. Additionally, the adiabatic expansion term depends on the local density, because gravity slows the expansion of overdense regions and hence decreases the cooling rate (and of course in underdense regions, the cooling accelerates).  Thus, the IGM will be seeded by small temperature fluctuations reflecting its density structure.

To describe these fluctuations, we write $\delta_T$ as the fractional gas temperature fluctuation and $\delta_\gamma$ as the photon density fluctuation (so that $\delta_\gamma = 4 \delta_{T_\gamma}$, the fractional CMB temperature fluctuation). Then the perturbed version of equation~(\ref{eq:Compton}) is 
\begin{equation}
{d \delta_T \over dt} = {2 \over 3} {d \delta_b \over d t} + {x_e(t) \over t_{\rm C}(z)} \left[ \delta_\gamma \left( {\bar{T}_\gamma \over \bar{T}_e} - 1 \right) + {\bar{T}_\gamma \over \bar{T}_e}
(\delta_{T_\gamma} - \delta_T) \right],
\label{eq:deltaT-fluc}
\end{equation} 
where the first term describes adiabatic cooling due to expansion (allowing for variations in the expansion rate) and the second accounts for variations in the rate of energy exchange through Compton scattering (which can result from variations in either the gas or photon temperatures).

Meanwhile, the fluctuations in the baryon temperature affect the baryon density evolution as well.  Equation~(\ref{eq:linper}) implicitly assumed that temperature fluctuations were driven (only) by density fluctuations; allowing a more general relation, we obtain \cite{naoz05}
\begin{equation}
\frac{\partial^2\delta}{\partial t^2}+2 H
\frac{\partial\delta}{\partial t}={3 \over 2} H^2 \left( {\Omega_c
\delta_c + \Omega_b \delta_b} \right) - {k^2 \over a^2} {k_B \bar{T}_e
\over \mu m_H} (\delta_b + \delta_T).
\label{eq:linper-Tfluc}
\end{equation}
 This, together with equations~(\ref{eq:deltaT-fluc}), (\ref{eq:Compton}), and a more precise version of~(\ref{eq:dxedt}) for the temperature and ionized fraction evolution, provide a complete set of equations to trace the density and temperature evolution (modulo one more effect that we will discuss next).  
 
Figure~\ref{fig:nb-pk} shows the resulting power spectra for the dark matter density, baryon density, baryon temperature, and photon temperature perturbations at four different redshifts.  The photon fluctuations are not directly observable, but the others can in principle be probed through the 21-cm line.  The photon perturbations are strongly suppressed on scales below the sound horizon thanks to their large pressure.  Near recombination, the baryonic perturbations are also suppressed on these scales, especially in the temperature, because they interact so strongly with the CMB.  However, after recombination, the baryons fall into the dark matter potential wells,
with their perturbations rapidly growing, and temperature fluctuations also grow thanks largely to the variations in the adiabatic cooling rate.  The turnover at very small scales in the baryonic power spectrum is due to the finite pressure of the gas. 

\subsubsection{Relative Streaming of Baryons and Cold Dark Matter} 

There is one additional effect on the baryonic power spectrum that may provide insight into cosmology during this era: ``streaming" of baryonic matter relative to dark matter \cite{tseliakhovich10}. As a relativistic fluid, CMB photons have a very high pressure that drives acoustic waves throughout that component. While these photons are coupled to the baryons through Compton scattering, they can drag the baryonic component along with them -- it is these acoustic waves in the photon-baryon fluid that we see in CMB fluctuations. Once recombination occurs, the radiation drag force decreases, and the baryons begin to fall into the potential wells of dark matter fluctuations (which have not participated in the acoustic waves). This transition can be seen in the dotted and short-dashed curves in Figure~\ref{fig:nb-pk}. Because the radiation sound speed is $\sim c/\sqrt{3}$, it is near the causal horizon at the time of recombination, corresponding to $\sim 150$~comoving Mpc today, where they can be observed as ``baryon acoustic oscillations" in the matter power spectrum (and, because their physical scale is well-known, used as a standard ruler to measure cosmological parameters). 

\begin{figure}
\centerline{\includegraphics[width=10cm]{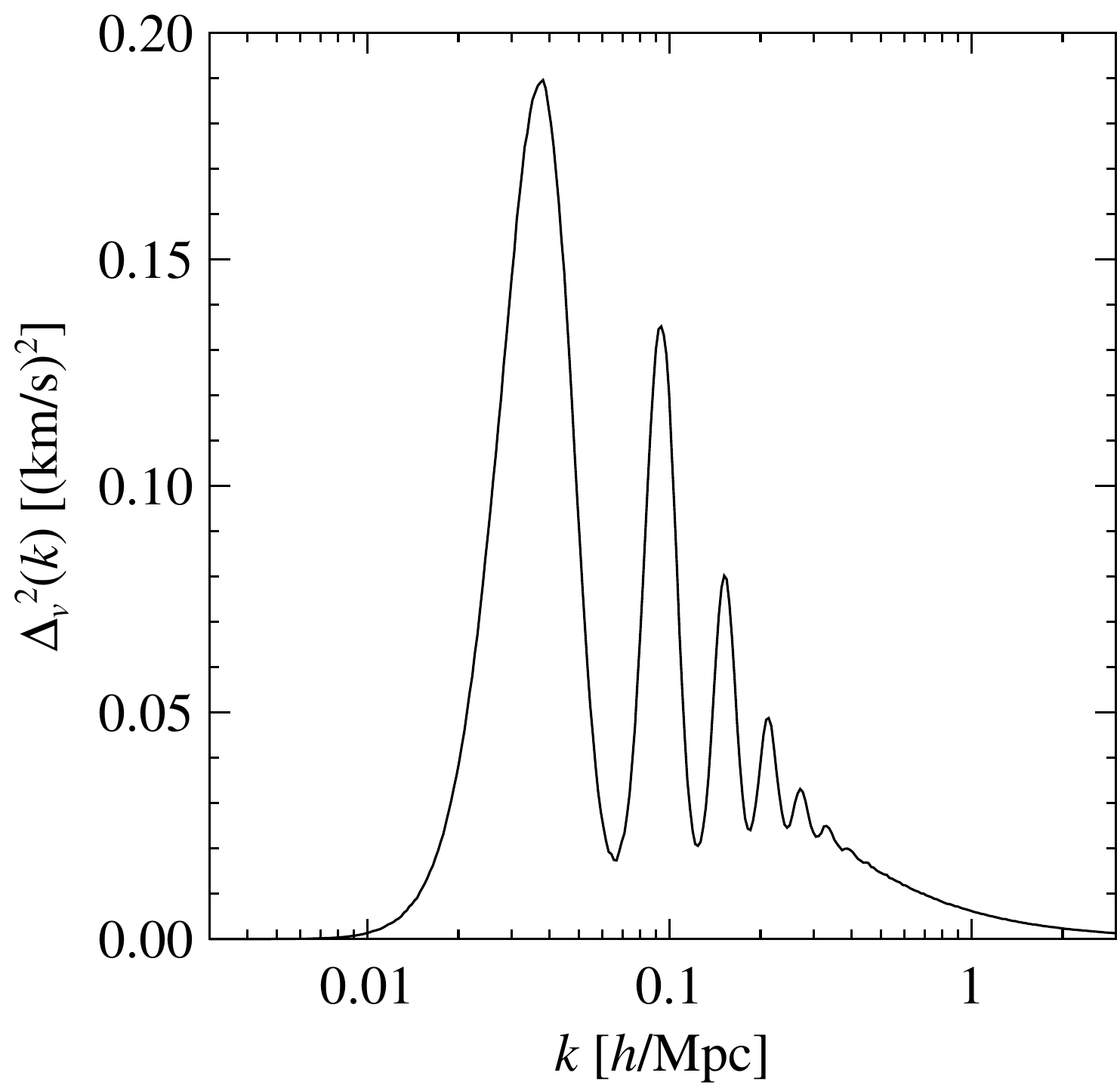}}
\caption{The power spectrum of the velocity difference perturbations 
(in units of between baryons and dark matter as a
function of comoving wavenumber $k$ at $z=15$.  Reproduced from N. Dalal, U.-L. Pen, \& U. Seljak. {\it Journal of Cosmology \& Astroparticle Physics} {\bf 2010}, 007. Copyright 2010 IOP Publishing and Sissa Medialab srl. All rights reserved. }
\label{fig:delta_v}
\end{figure}

There is an additional effect relevant to high redshifts, however. Even after the radiation driving becomes ineffective, the baryons are left with a relic velocity imprinted by the pressure force with a root-mean-square ({\it rms}) speed of $v_{\rm bc}\approx 10^{-4}c=30~{\rm km~s^{-1}}$, which decays with redshift as
$1/a$ \cite{tseliakhovich10}. This \emph{streaming velocity} between the baryonic and dark matter components is coherent over very large scales -- comparable to the acoustic scale -- and fades only slowly over time. Figure~\ref{fig:delta_v} shows the variance of the relative velocity perturbations  as a function of the mode wavenumber $k$ at $z=15$. The power is substantial on scales up to the sound horizon at decoupling ($\sim 140$ comoving Mpc), but it declines rapidly at $k>0.5~{\rm Mpc^{-1}}$, indicating that the streaming velocity is coherent over a scale of several comoving Mpc. Therefore, in the rest-frame of small-scale fluctuations (such as those that will eventually collapse into galaxies), the baryons appeared to be moving coherently as a ``wind," which can suppress the formation of the first galaxies \cite{dalal10, fialkov14}.

\subsection{The Global 21-cm Signal During the Dark Ages} \label{cos-21cm-dark}

\begin{figure}
\centerline{\includegraphics[width=8cm]{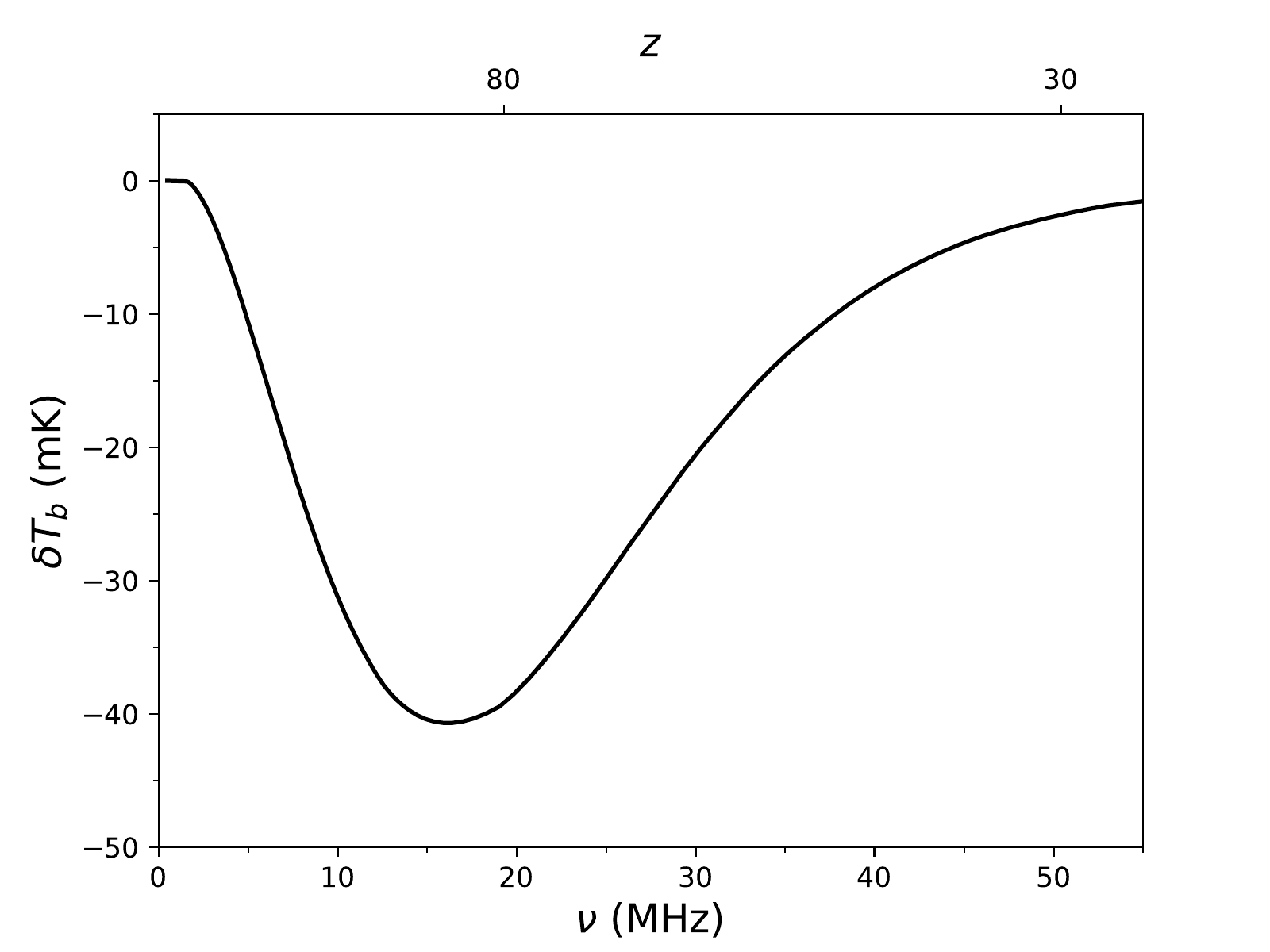} \includegraphics[width=8cm]{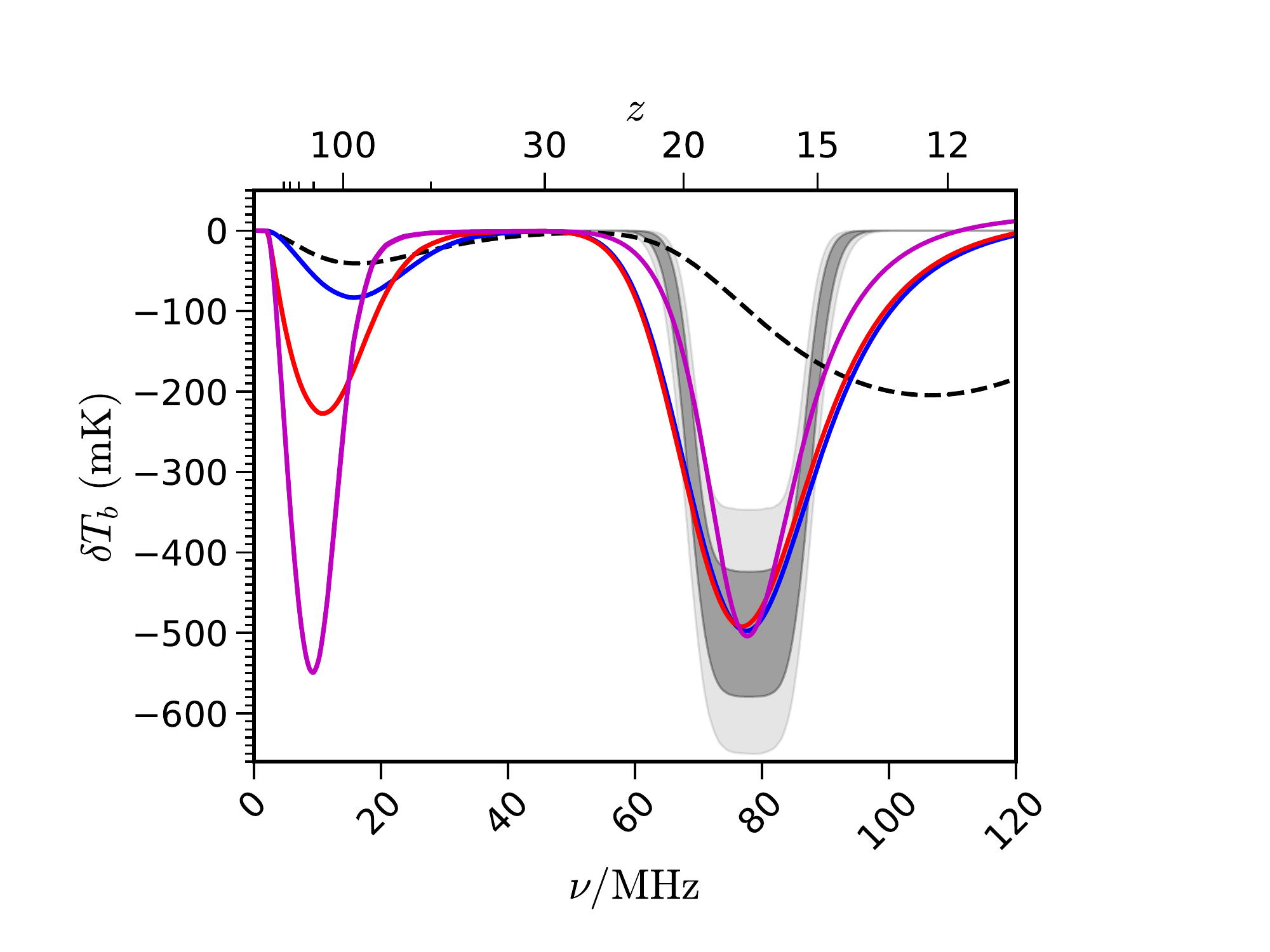}}
    \caption{\emph{Left:} The Dark Ages 21-cm absorption trough in the standard cosmology. The shape at $z > 30$ is independent of astrophysical sources. \emph{Right:} The black dashed line shows the mean 21-cm brightness temperature (averaged across the sky) in the standard cosmology. The gray contours show schematically the reported EDGES absorption signal \cite{Bowman18}. The solid curves are phenomenological models that invoke extra cooling to match the amplitude of the EDGES signal but that also dramatically affect the Dark Ages absorption trough at $z>50$. Courtesy J. Mirocha, based on calculations in \cite{mirocha19}.}
    \label{fig:sample-histories}
\end{figure}

With the thermal history from Figure~\ref{fig:darkage-thermal}, we can easily calculate the global 21-cm signal throughout the Dark Ages -- without any way of generating a Lyman-$\alpha$ background, the competition between the CMB and particle collisions sets the spin temperature. The left panel of Figure~\ref{fig:sample-histories} shows the result: once the gas temperature begins to fall below the CMB temperature at $z \sim 150$, the 21-cm signal can be seen in absorption. At these early times, the gas is relatively dense and has a high enough temperature for collisional coupling to be substantial, so $T_S \rightarrow T_K$. However, as the gas expands and cools, collisional coupling becomes inefficient, and eventually the spin temperature begins to return to $T_\gamma$. By the time we expect star formation to begin in earnest ($z < 30$), $T_S \approx T_\gamma$, so the 21-cm line is nearly invisible. 

The global 21-cm signal is therefore sensitive to the IGM thermal history, so any process that affects the temperature evolution can also be probed by the 21-cm line. This method provides a particularly powerful probe of non-standard physics because the low gas temperature during this period makes the 21-cm line a sensitive calorimeter of additional heating (or cooling) and/or of an excess radio background over and above the CMB \cite{feng18}.

While many such processes tend to \emph{heat} the IGM and therefore \emph{decrease} the amplitude of the 21-cm signal (e.g. \cite{chen04-decay, furl06-dm,shchekinov07, chuzhoy08}), the recent claim of a detection of a 21-cm absorption feature at 78~MHz ($z \sim 17$) by the EDGES collaboration \cite{Bowman18} has triggered interest in non-standard models that \emph{amplify} the 21-cm signal (though note that the EDGES signal has not yet been independently confirmed, and with such a challenging analysis systematics and foreground contaimnation remain a concern \cite{hills18, draine18, spinelli19, bradley19}). The EDGES feature is more than twice as deep as expected if gas follows the standard thermal history, even if astrophysical sources were able to drive $T_S \rightarrow T_K$ and no X-ray heating had yet occurred. It therefore requires new physics that either \emph{cools} the IGM or \emph{increases} the radio background against which the gas absorbs. A multitude of processes has been proposed to explain the anomalous detection, including dark matter-baryon scattering \cite{munoz15, barkana18, slatyer18, hirano18}, millicharged dark matter \cite{munoz18-cool, berlin18, kovetz18}, axions \cite{moroi18}, neutrino decay \cite{chianese19}, charge sequestration \cite{falkowski18}, quark nuggets \cite{lawson19}, dark photons \cite{jia19}, and interacting dark energy \cite{costa18}, and it has been used to constrain additional exotic processes like dark matter annihilation \cite{cheung19}.

Although the EDGES signal at $z \sim 17$ is very likely past the end of the Dark Ages -- after the first astrophysical sources formed -- the same new physics would have affected the Dark Ages signal. The solid curves in Figure~\ref{fig:sample-histories} illustrate how models that invoke excess cooling to explain the EDGES result (shown schematically by the gray contours) could also greatly amplify the Dark Ages signal. (Here the curves use a phenomenological parameterized cooling model as in \cite{mirocha19}; physically-motivated models will differ in the details.) (Note that the signal still vanishes at $z \sim 30$ because collisional coupling is still inefficient.) In these cases, even though the exotic physics have implications at relatively low redshifts, Dark Ages observations serve to break degeneracies between astrophysics and that new physics.

\subsection{The Power Spectrum During the Dark Ages} \label{cos-21cm-ps}

\begin{figure}
	\centerline{\includegraphics[width=7.5cm]{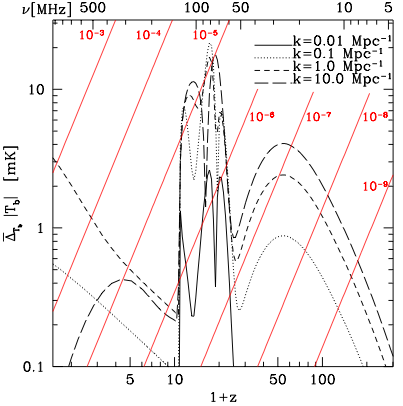}}
    \caption{ 21-cm fluctuations are substantial during the Dark Ages. The curves show the amplitude of the 21-cm brightness temperature fluctuations at several different wavenumbers from the Dark Ages to low redshifts, in the standard model of cosmology. Red diagonal lines compare these fluctuations to the foreground brightness temperature (from Galactic synchrotron): each scales the foreground by the number shown. Note that exotic cooling scenarios described in \S \ref{cos-21cm-dark} could significantly amplify these fluctuations. From \cite{pritchard12}. Copyright IOP Publishing. Reproduced with permission. All rights reserved.}
    \label{fig:pritchard-flucs}
\end{figure}

As we have discussed, maps of 21-cm emission provide a sensitive probe of the power spectrum of density and temperature fluctuations \cite{kleban07, mao08}. Figure~\ref{fig:pritchard-flucs}, from \cite{pritchard12}, shows examples of how modes of the 21-cm fluctuation power spectrum evolve in the standard cosmology. Although the strongest fluctuations occur at $z \sim 10$, when astrophysical sources dominate, the Dark Ages signal are not too far behind: this is because, although the fractional density fluctuations are small at those times, the mean temperature can be relatively large during the absorption era -- even in the standard cosmology.

The 21-cm power spectrum during the Dark Ages offers several advantages over other probes of the density field. Because they use the cosmological redshift to establish the distance to each observed patch, 21-cm measurements probe three-dimensional volumes --- unlike the CMB, which probes only a narrow spherical shell around recombination. Additionally, the 21-cm line does not suffer from Silk damping (photon diffusion), which suppresses the CMB fluctuations on relatively large scales. The number of independent modes accessible through this probe is therefore \cite{loeb04}
\begin{equation}
N_{\rm 21cm} \sim 8 \times 10^{11} \left( {k_{\rm max} \over 3 \ \mbox{Mpc}^{-1}} \right)^3 \left( {\Delta \nu \over \nu} \right) \left( {1+z \over 100} \right)^{-1/2},
\label{eq:nmax}
\end{equation}
where $\Delta \nu$ is the bandwidth of the observation. The choice of $k_{\rm max}$ --- the smallest physical scale to be probed --- is not obvious. The Jeans length during the Dark Ages corresponds to $k_{\rm max} \sim 1000\ \rm{Mpc}^{-1}$. Accessing these small-scale modes in three dimensions would require an enormous instrument, but our relatively conservative choice in equation~(\ref{eq:nmax}) shows that even a more modest effort provides a massive improvement over the information contained in all the measurable modes of the CMB, $N_{\rm CMB} \sim 10^7$. Moreover, because the density fluctuations are still small at such early epochs, these modes remain in the linear or mildly non-linear regime, allowing a straightforward interpretation of them in terms of the fundamental parameters of our Universe \cite{lewis07}.  

\begin{figure}
	\centerline{\includegraphics[width=9cm, clip=True, trim=0cm 0.8cm 0cm 0cm]{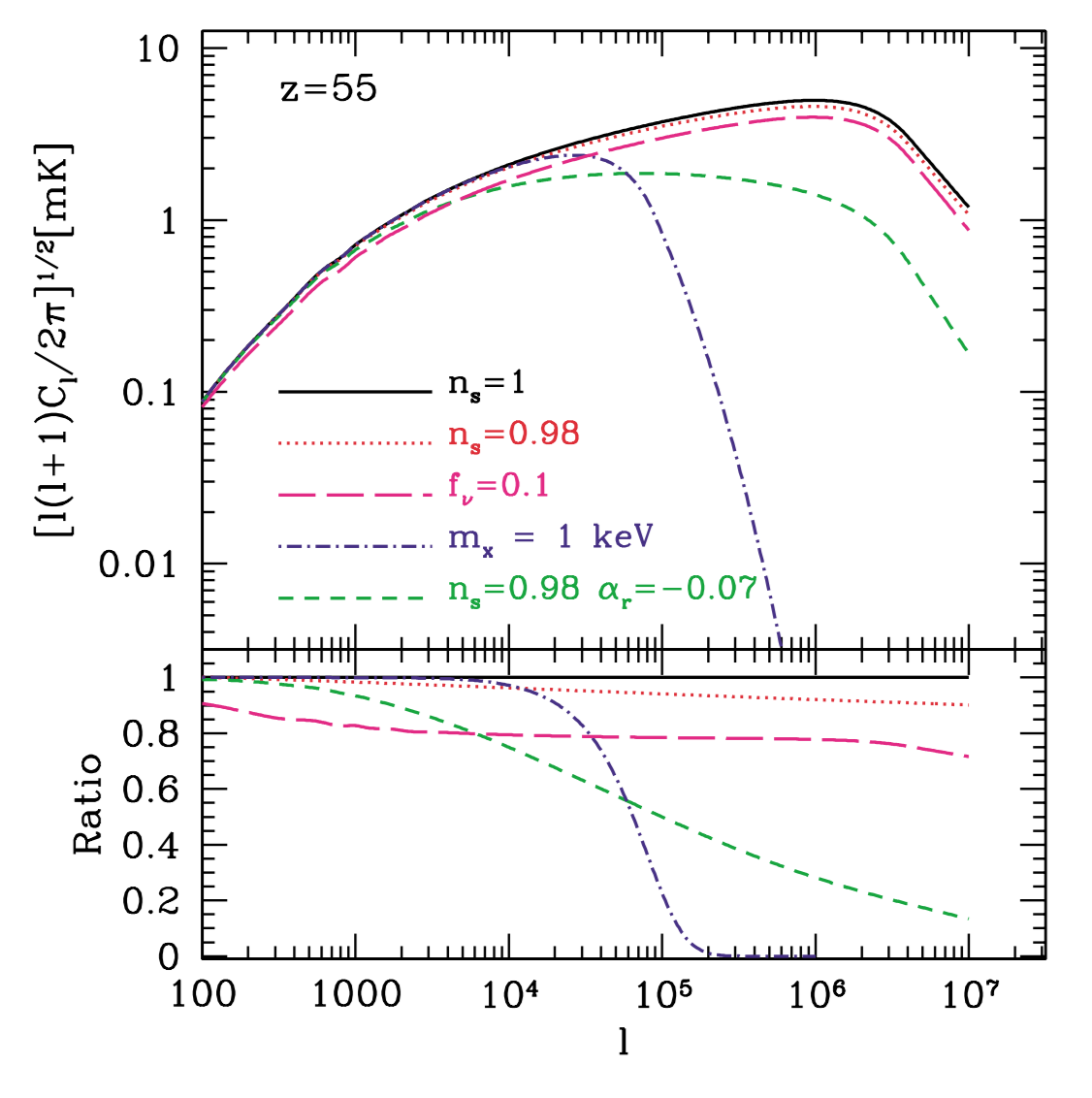}}
    \caption{The 21-cm power spectrum is a sensitive probe of cosmological parameters. The angular power spectrum of 21-cm fluctuations at $z=55$ in ``standard" cosmologies, varying some of the parameters. The solid and dotted curves use $\Lambda$CDM with the power law index of density fluctuations $n_s=1$ and 0.98; respectively; the short-dashed curve adds a ``running" to the spectral index. The long-dashed curve assumes 10\% of the matter density is in the form of massive neutrinos (with masses 0.4~keV), while the dot-dashed curve assumes warm dark matter with particle masses of 1~keV.  Reprinted figure from Loeb, A. \& Zaldarriaga, M., {\it Phys. Rev. Lett.} {\bf 92}, 21, 211301. Copyright 2004 by the American Physical Society.}
    \label{fig:sample-cos}
\end{figure}

In principle, measurements of the 21-cm power spectrum during the Dark Ages will therefore enable a number of precision cosmological measurements, even within the ``standard" cosmology. For example, because 21-cm fluctuations extend to such small scales, they expand the dynamic range of power spectrum measurements over several orders of magnitude. This is useful for a number of specific cosmological parameters, as illustrated by the curves in Figure~\ref{fig:sample-cos}, but also including the running of the spectral index of the matter power spectrum \cite{mao08}, a key parameter that can test the underlying assumptions of the inflationary paradigm, as well as the total spatial curvature and neutrino masses \cite{mao08}.

The large number of modes is also essential to constraining primordial non-Gaussianity in the cosmological density field, which is another tool for testing  inflationary models. CMB measurements already offer constraints, which can best be improved by larger-volume surveys. However, low-$z$ galaxy surveys suffer from contamination, because nonlinear structure formation also generates non-Gaussianity. The ``clean," small-scale Dark Ages 21-cm field is an excellent opportunity to further constrain the non-Gaussianity  \cite{chen18} -- in principle, 21-cm Dark Ages measurements can test the generic inflationary picture itself \cite{munoz15-ng}. 

The potential insights we have discussed so far are useful within the standard cosmology, but the exotic physics mechanisms discussed in the previous section will inevitably affect the 21-cm power spectrum as well, though the implications have only recently begun to be explored. Most obviously, the power spectrum amplitude is proportional to the square of the mean brightness temperature. But more subtle signatures that depend on the particulars of the physics can also appear. For example, if the cooling is triggered by scattering between the baryons and a fraction of the dark matter that has a modest charge, the scattering rate will be modulated by the relative streaming velocity of dark matter and baryons. Such a model therefore leaves distinct features in the 21-cm power spectrum that trace the velocity structure \cite{munoz19, munoz19-ruler}. The detailed implications of most of these exotic models are mostly unexplored, but any model in which (1) the energy exchange depends on local density, velocity, or temperature; or (2) in which background radio energy deposition is inhomogeneous, should generically leave signatures in the power spectrum. 

\section{Cosmology During the Era of Astrophysics} \label{cos-astro}

Although the Dark Ages offer a clean and powerful testbed to probe cosmology, in practice they will be difficult to explore. Not only do the astrophysical foregrounds become significantly stronger at low frequencies (as shown by the diagonal lines in Figure~\ref{fig:pritchard-flucs}, which increase toward higher redshifts), but Earth's ionosphere also becomes opaque at very low frequencies \cite{datta14}, which may necessitate observations from space or the Moon.

Because the 21-cm background will likely be observed at lower redshifts -- after the first luminous sources have appeared -- we will next consider how cosmological information might be extracted from observations during the Cosmic Dawn. The effects we have already described will also affect the global 21-cm signal and the power spectrum during the later era. However, many astrophysical mechanisms (those described in Chapter~2) also affect the 21-cm background during this era, so the challenge is to separate the cosmological information from the astrophysical. In this section, we shall consider some strategies to do that.

\subsection{Isolating the Matter Power Spectrum} \label{cos-ps}

Conceptually, the simplest method is simply to measure the 21-cm power spectrum and fit simultaneously for both astrophysical and cosmological parameters. We will discuss the practicalities of such fits in Chapter~4, but it will suffice for now to note that it is not a trivial exercise, and adding additional cosmological parameters will be a challenge \cite{clesse12, kern17, park19}. Moreover, the astrophysical effects can be very strong and hence mask any cosmological effects. 

Short of modeling the astrophysics precisely enough to extract cosmology -- a strategy whose potential won't really be known until the astrophysics is better understood\footnote{However, note that recent work has shown that the ionization field may be constructed as a perturbative expansion around the density field in certain regimes, which suggests precision modeling may indeed be possible \cite{mcquinn18, hoffmann19}.} -- the chief hope for isolating the power spectrum is that there exists an era in which astrophysics processes can largely be ignored. This is not an entirely unreasonable expectation. Recall from Chapter~1 that the 21-cm brightness temperature is
\begin{equation}
T_b(\nu) \approx  9\;x_{\rm HI}(1+\delta) \, (1+z)^{1/2}\, \left[1-\frac{T_{\gamma}(z)}{T_S}\right] \, \left[ \frac{H(z)/(1+z)}{d v_\parallel/d r_\parallel} \right] \ \mbox{mK}.
\label{eq:Tb-cos}
\end{equation}
Astrophysical effects determine the neutral fraction $x_{\rm HI}$ and the spin temperature $T_S$, but the other factors -- density and velocity -- are driven by cosmological processes. Thus we can imagine that cosmological information will show up clearly in the power spectrum if, for example, there exists a period in which $T_S \gg T_\gamma$ (so that the temperature effects can be ignored) but in which ionization fluctuations are not yet significant. 

Simple estimates show that such a period is far from impossible -- but also not guaranteed. Reionization requires at least one ionizing photon per baryon, or of order $\sim 10$~eV of ionizing energy released by stars per baryon. Heating the IGM above the CMB temperature -- so that $(1 - T_\gamma/T_S) \approx 1$ requires only $\sim 10^{-2}$~eV (corresponding to a temperature $\sim 100$~K, though only a fraction of the X-ray energy would actually be used to heat the IGM; see \S \ref{xrayheat}). Thus if early sources produce at least $\sim 10^{-3}$ as much energy in X-rays as they do in ionizing photons, heating would occur before reionization -- leaving open the possibility that a period exists in which astrophysics can be ignored. More complex astrophysical processes can also enable such a period as well -- for example, strong photoheating feedback can delay reionization relative to X-ray heating \cite{mesinger13}. Whether this is more than speculation remains uncertain: calibrating the X-ray luminosity of star-forming galaxies to local measurements suggests that the reionization epoch may overlap with the X-ray heating epoch \cite{mirocha17, park19}, but if the EDGES measurement is confirmed, heating must actually occur very early in the Cosmic Dawn, well before reionization is complete.

\subsection{Redshift Space Distortions} \label{cos-redshift-space}

To this point, we have largely ignored the last factor in equation~(\ref{eq:Tb-cos}): the velocity gradient. However, it offers another route to extracting cosmological information. Usually, we expect the fluctuations from the other terms -- density, ionization fraction, Ly$\alpha$ flux, and temperature -- to be  isotropic, because the processes responsible for them have no preferred direction [e.g., $\delta({\bf k}) = \delta(k)$]. However, peculiar velocity
gradients introduce anisotropic distortions through the ``Kaiser effect" \cite{kaiser84}, which emerge because 21-cm observations use the line's observed frequency as a proxy for the distance of the cloud that produced it. 

Consider a spherical overdense region. Because of its enhanced gravitational potential, the region expands less quickly than an average region of the same mass. Therefore, to an observer using redshift as a distance indicator, the apparent \emph{radial} size of the overdense region is smaller than that of the average region. Of course, its transverse size can be measured by its angular extent on the sky so is unaffected by the velocity structure. Thus, to the observer, the spherical region is distorted, appearing larger along the plane of the sky than in the radial direction. An underdense region is distorted in the opposite way: because it expands faster than average, it appears larger in the radial direction than along the plane of the sky. Thus redshift space distortions will \emph{exaggerate} intrinsic density fluctuations, but they do so in an anisotropic way -- making this source of fluctuations separable from others \cite{barkana05-vel, mcquinn06-param}.

To see these effects, we start by labeling the coordinates in redshift space with ${\bf s}$. Assuming that the radial extent of the volume is small, so that the Hubble parameter $H$ is constant throughout the volume, these coordinates are related to the real space ${\bf r}$ by 
\begin{equation} 
{\bf s}({\bf r}) = {\bf r} + {U({\bf r}) \over H}, 
\end{equation} 
where $U({\bf r}) = {\bf v} \cdot \hat{\bf x}$ is the radial component of the peculiar velocity.

Now consider a set of particles with number density $n({\bf r})$ that are biased with respect to the dark matter by a factor $b$. Number conservation demands that the fractional overdensity in redshift space is related to that in real space via $[1 + \delta_s({\bf s})] d^3 {\bf s} = [1 + \delta({\bf r})] d^3 {\bf r}$. The Jacobian of the transformation is \begin{equation}
d^3{\bf s} = d^3{\bf r} \left[ 1 + {U({\bf r}) \over r} \right]^2 \left[ 1 + {dU({\bf r}) \over dr} \right], 
\end{equation} 
because only the radial component of the volume element, $r^2 dr$, changes from real to redshift space.  Thus the density observed in redshift space increases if the peculiar velocity gradient is smaller than the Hubble flow, while the redshift space density will be smaller if the peculiar velocity gradient is larger.  Thus, assuming $|U(r)| \ll Hr$, 
\begin{equation}
\delta_s({\bf r}) = \delta({\bf r}) - \left( {d \over dr} + {2 \over r} \right) {U(r) \over H}.
\label{eq:delta-s}
\end{equation}
Importantly, the peculiar velocity field itself is a function of the dark matter density field, as described qualitatively above. More rigorously, in Fourier space the components of the peculiar velocity ${\bf u}_{\bf k}$ are directly related to those of the density field, because the latter sources the gravitational fluctuations that drive the velocity gradients (e.g., \cite{kaiser84}):
\begin{equation} 
{\bf u}_{{\bf k}} = -i {a H f(\Omega) \over k} \delta_{\bf k} \hat{\bf k},
\label{eq:vpec}
\end{equation} 
where $f(\Omega)$ is a function of the growth rate of cosmological perturbations.  

Equation~(\ref{eq:delta-s}) shows that there are two corrections from the redshift space conversion. To see which of these dominates, consider a plane wave perturbation, $U \propto e^{i {\bf k} \cdot {\bf r}}$. Then the derivative term is $\sim kU/H_0$ while the last term is $\sim U/H_0 r$.  But $r$ is the median distance to the survey volume, and $k$ corresponds to a mode entirely contained inside it, so $k r \gg 1$, and we may neglect the last term.  If we further make the small-angle approximation and make a Fourier transform, so that $\hat{\bf x}$ is also approximately a constant over the relevant volume, the Fourier transform of equation~(\ref{eq:delta-s}) is 
\begin{equation} 
\delta_s({\bf k}) = \delta({\bf k})[ 1 + \beta \mu_{\bf k}^2], 
\end{equation} 
where $\mu_{\bf k} = \hat{\bf k} \cdot \hat{\bf x}$ is the cosine of the angle between the wave vector and the line of sight. Here $\beta = f(\Omega_m)/b$ corrects for a possible bias between the tracers we are studying and the growth rate of dark matter perturbations.  For the case of 21-cm fluctuations in the IGM gas, the bias factor is very close to unity except below the Jeans filtering scale.  

The redshift-space distortions therefore provide an anisotropic \emph{amplification} to the background signal, because only modes along the line of sight are affected. Averaged over all modes, these distortions amplify the signal by a factor $\approx \VEV{(1+ \mu^2)^2} \approx 1.87$ \cite{bharadwaj04-vel}.

For the purposes of extracting cosmological information, the anisotropies are helpful in that they imprint angular structure on the signal, which may allow us to separate the many contributions to the total power spectrum \cite{barkana05-vel}. Brightness temperature fluctuations in Fourier space have the form
\begin{equation}
\delta_{21} =  \mu^{2} \beta \delta + \delta_{\rm iso} 
\label{eq:d21ft}
\end{equation}
where we have collected all the statistically isotropic terms -- including those due to astrophysics -- into $\delta_{\rm iso}$. Neglecting ``second-order" terms (see below) and setting $\beta=1$, the total power spectrum can be written  
\begin{equation}
P_{21}({\bf k}) = \mu^{4} P_{{\delta} {\delta}} + 2 \mu^{2} P_{{\delta}_{\rm iso} {\delta}} + P_{{\delta}_{\rm iso} {\delta}_{\rm iso}}.
\label{eqn:p_polynomial}
\end{equation}
(Here we have written the normal density power spectrum as $P_{\delta \delta}$ for clarity.) By separately measuring these three angular components (which requires, in principle, estimates at just a few values of $\mu$), we can isolate the contribution from density fluctuations $P_{\delta \delta}$. Measuring this component, without any astrophysical contributions, will provide the desired cosmological constraints. 

However, in practice the angular dependence of the power spectrum will not be so simple. To write equation~(\ref{eqn:p_polynomial}), we must neglect ``second-order" terms in the perturbation expansion of the 21-cm field, such as the density and the ionization field perturbations. But the latter is not actually a small term (because, at least in the standard reionization scenarios, $x_{\rm HI}=0$ or 1), so its contributions do not decrease rapidly in higher-order terms \cite{lidz07-ng}. During reionization, these additional terms complicate the angular dependence and will significantly complicate attempts to separate the $\mu^n$ powers during reionization more difficult \cite{mcquinn06-param, shapiro13}. Moreover, if the first H~II regions are highly biased -- thus overlapping the regions with the largest peculiar velocities -- the redshift space distortions can be suppressed \cite{mesinger11, mao12}. The redshift space distortions are also more complicated if the heating and reionization eras overlap \cite{ghara15}.
 
\begin{figure}
	\centerline{\includegraphics[width=9cm, clip=True, trim=0cm 0.6cm 0cm 0.5cm]{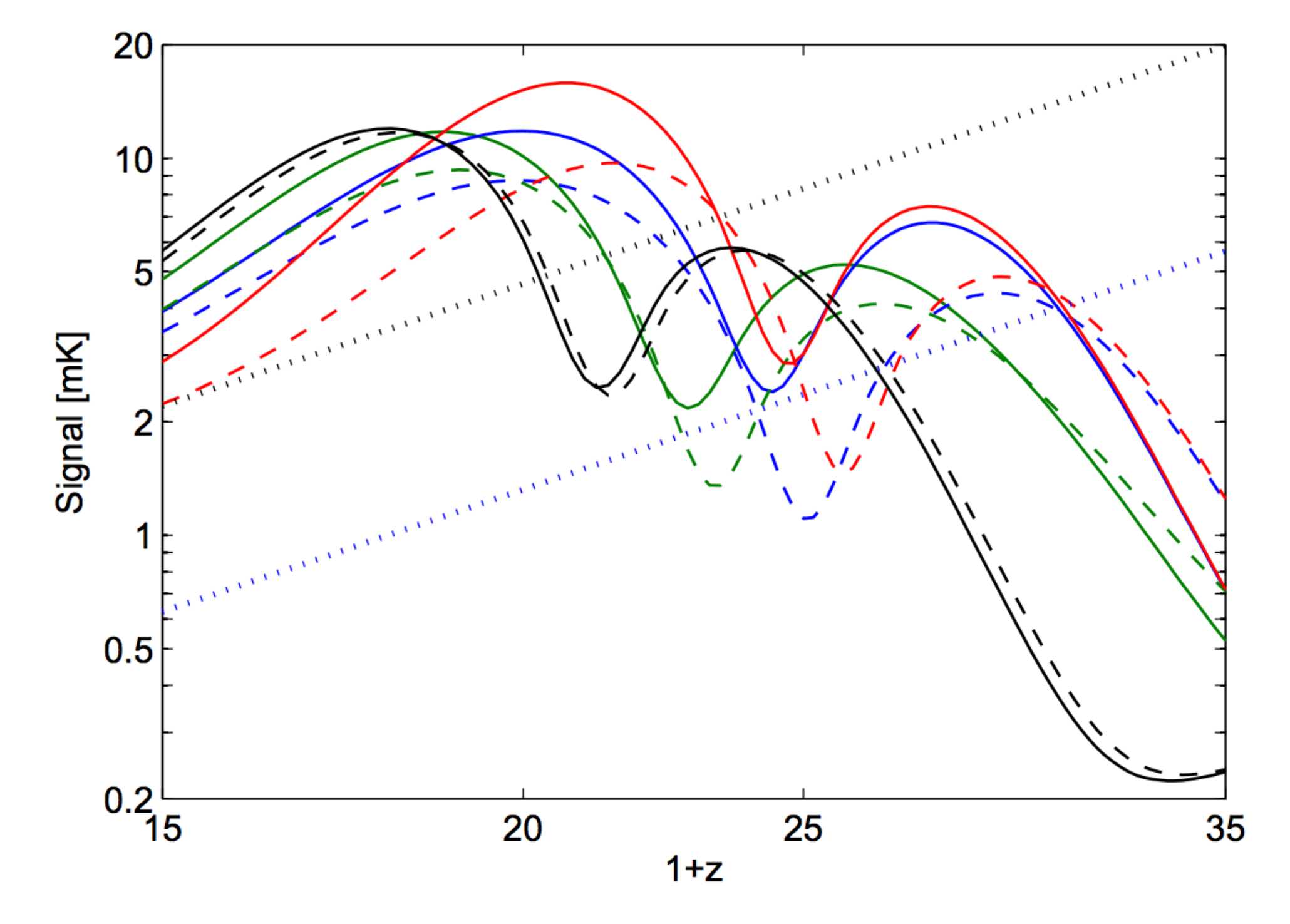}}
    \caption{The relative streaming velocity between baryons and dark matter can impact the 21-cm power spectrum during the Cosmic Dawn. The solid curves show the time evolution of several example 21-cm power spectra, evaluated at $k \sim 0.1 h$~Mpc$^{-1}$, without including streaming, while the dashed curves show the same scenarios with streaming included. The dotted lines show expected sensitivities of current and future 21-cm radio arrays. Reproduced from Fialkov, A. et al, ``Complete history of the observable 21 cm signal from the first stars during the pre-reionization era," \emph{Monthly Notices of the Royal Astronomical Society}, 437, L36-40. Copyright OUP 2014. }
    \label{fig:streaming}
\end{figure}

\subsection{Indirect Effects of Cosmology on the 21-cm Background} \label{cos-indirect}

Finally, it is also worth noting that cosmological processes can have direct effects on astrophysical sources and hence indirect effects on the 21-cm background. For example, consider warm dark matter. If dark matter has a non-zero velocity dispersion, then it can easily escape from shallow potential wells, suppressing the formation of small dark matter haloes. Because the first phases of galaxy formation occur in small haloes, warm dark matter delays galaxy formation, which in turn delays the formation of Lyman-$\alpha$, X-ray, and ionizing backgrounds and changes the timing (and potentially spatial fluctuations) of the 21-cm background (e.g., \cite{barkana01, yue12,lopezhonorez17}). Of course, this dark matter effect is degenerate with astrophysical processes (for example, strong feedback in small galaxies may also suppress their star formation rates) so requires careful analysis, and other cosmological changes can have qualitatively similar effects (e.g., \cite{yoshida03}), although such effects may be distinct from large swathes of astrophysical parameter space \cite{sitwell14}.

Another example is the interaction between astrophysical processes and the relative streaming between baryons and dark matter. The streaming velocity also suppresses the formation of small baryonic haloes, because a shallow potential well cannot accrete gas traveling by it at sufficiently large velocity. However, unlike in the case of warm dark matter, the streaming effect is spatially variable, so, at least in some circumstances, the streaming effect will imprint spatial structure on the radiation backgrounds and hence on the 21-cm power spectrum \cite{dalal10, fialkov14, munoz19}. Figure~\ref{fig:streaming} shows an example of this effect during the era in which the first stars appear.

Finally, dark matter annihilation offers an other interesting example of the interaction between cosmology and astrophysics. The heating from dark matter annihilation occurs very uniformly. If astrophysically-driven X-ray heating begins within such a pre-heated medium, the associated large-scale peak in the power spectrum occurs in emission rather than absorption, providing a distinct signature for (some) dark matter annihilation scenarios  \cite{evoli14, lopez16}. 

\begin{figure}
	\centerline{\includegraphics[width=9cm, clip=True, trim=0cm 0cm 0cm 0cm]{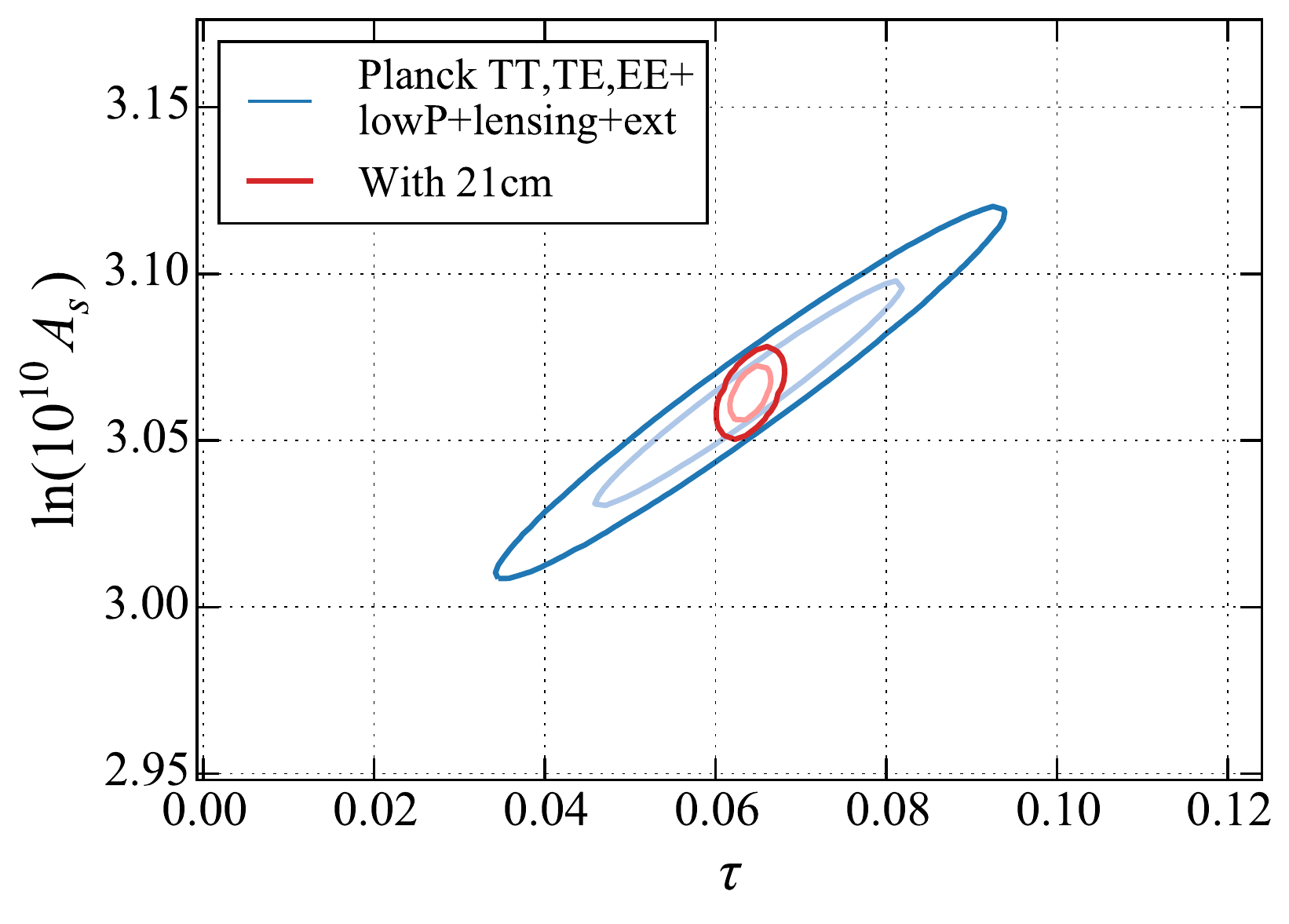}}
    \caption{Constraints on the amplitude of the primordial power spectrum, $A_s$, and the CMB optical depth to electron scattering, $\tau$. The elongated curves show the current constraints (at 1 and $2\sigma$) from combining the Planck satellite and several other probes. The tighter constraints show the combination with forecasted measurements from HERA, illustrating how the 21-cm measurement of the ionization history substantially improves constraints on other cosmological parameters. Figure based on calculations from \cite{liu16}, provided courtesy A. Liu.}
    \label{fig:cmb}
\end{figure}

\section{21-cm Cosmology in a Larger Context} \label{cos-complementary}

Although the focus of this book is on the 21-cm line itself, it is worth emphasizing that observations of the high-$z$ spin-flip background will ultimately be combined with many other observations. For cosmological measurements, it is therefore useful to understand synergies between the 21-cm background and other probes \cite{liu16, liu16-cmb}. Some of the key advantages of the 21-cm line have already been described: it can probe small physical scales and high-redshifts. Additionally, it can break degeneracies within other probes. Figure~\ref{fig:cmb} shows an example. If the 21-cm line can measure the reionization history, it provides an independent estimate of the CMB optical depth to electron scattering, which is otherwise nearly degenerate with the amplitude of the initial power spectrum -- and thus allow a precision measurement of that parameter.


\bibliographystyle{plain}
\bibliography{ch3-refs}

\end{document}